\documentclass[useAMS,usenatbib]{mn2e}
\usepackage{amsmath,amssymb,epsfig,natbib,bm,psfrag,ifthen,url}
\usepackage{hyperref}
\usepackage{epsfig}
\usepackage{graphicx}
\usepackage{psfrag}
\usepackage{url}
\usepackage{mathrsfs}
\usepackage[normalem]{ulem}

\usepackage{amsmath,amssymb}

\def\eprinttmp@#1arXiv:#2 [#3]#4@{
\ifthenelse{\equal{#3}{x}}{\href{http://arxiv.org/abs/#1}{#1}
}{\href{http://arxiv.org/abs/#2}{arXiv:#2} [#3]}}

\providecommand{\eprint}[1]{\eprinttmp@#1arXiv: [x]@}
\newcommand{\adsurl}[1]{\href{#1}{ADS}}

\usepackage{color}

\title[Optimising cosmic shear surveys to measure modifications to gravity]
{Optimising cosmic shear surveys to measure modifications to gravity on cosmic scales}
\author[Donnacha Kirk, Istvan Laszlo, Sarah Bridle, Rachel Bean]{Donnacha Kirk$^{1}$,Istvan Laszlo$^{2}$, Sarah Bridle$^{1}$, Rachel Bean$^{2}$\\
$^{1}$Department of Physics \& Astronomy, University College London, Gower Street, London, WC1E 6BT, UK\\
$^{2}$Department of Astronomy, Cornell University, Ithaca, NY 14853, USA}

\begin {document}

\pagerange{\pageref{firstpage}--\pageref{lastpage}} \pubyear{2009}

\maketitle

\label{firstpage}

\begin{abstract}
We consider how upcoming photometric large scale structure surveys can be optimized to measure the properties of dark energy and possible cosmic scale modifications to General Relativity in light of realistic astrophysical and instrumental systematic uncertainities.
In particular we include flexible descriptions of intrinsic alignments,
galaxy bias and photometric redshift uncertainties in
a Fisher Matrix analysis of shear, position and position-shear correlations,
including complementary cosmological constraints from the CMB.
We study the impact of survey tradeoffs in depth
versus breadth, and redshift quality. We parameterise the results in terms of the Dark Energy Task Force
figure of merit, and deviations from General Relativity through an analagous Modified Gravity
figure of merit. We find that intrinsic alignments weaken the dependence of
figure of merit on area and that, for a fixed observing time, halving the area of a Stage IV reduces the figure of merit by 20\% when IAs are not included and by only 10\% when IAs are included.
While reducing photometric redshift scatter improves constraining power, the dependence is shallow.
The variation in constraining power is stronger once IAs are included and is slightly more pronounced for MG constraints than for DE. The inclusion of intrinsic alignments and galaxy position information reduces the required prior on photometric redshift accuracy by an order of magnitude for both the fiducial Stage III and IV surveys, equivalent to a factor of 100 reduction in the number of spectroscopic galaxies required to calibrate the photometric sample.

\end{abstract}


\section{Introduction}
The dawn of ``precision cosmology'' was heralded by the results of surveys which, for the first time, produced data of sufficient quantity and quality that our cosmological probes could begin to accurately measure some of the fundamental properties of the Universe. The outcome has been the $\Lambda$CDM concordance cosmology, a description of the Universe compatible with the joint constraints from many sources of cosmological information.
The next decades will see a step-change in our ability to measure cosmological parameters as new surveys produce orders of magnitude more data than has previously been available. This will allow us to test the standard cosmological model as never before.

The standard $\Lambda$CDM model describes a Universe made up of $\sim$75\% Dark Energy, a smooth, negative pressure fluid, $\sim$20\% Dark Matter, collisionless massive particles which interact solely via gravity, and just $\sim$5\% baryons making up the
potentially visible mass of the Universe \citep{dunkleyetal}. The standard model also assumes gravity to be described by Einstein's General Relativity (GR).

A Universe governed by GR and populated by standard gravitating matter cannot explain the observed acceleration. The dark energy fluid was proposed to solve this paradox, reviving the idea of a cosmological constant which Einstein had originally included in his GR field equations and subsequently discarded. Much effort has been devoted to characterising the nature of dark energy- particularly attempts to discriminate between a pure cosmological constant and a dynamic scalar potential with time-varying equation of state $w(a)$ \citep{detf}. In all its forms, dark energy poses problems of fine-tuning for which we have, as yet, no physical motivation.

Rather than invoke dark energy, an alternative explanation for cosmic acceleration has been proposed -- that our standard theory of gravity is incomplete and that a correct theory of gravity would explain cosmic acceleration at late times and large scales in a universe populated by  matter with positive pressure. There are a large number of theoretically motivated modified theories of gravity, see \citet{jain_khoury} for a review.
Testing the theory of gravity on cosmic scales will be one of the most interesting opportunities afforded by upcoming survey data.

Weak gravitational lensing (WL)  is a particularly useful probe of gravity because it is sensitive to $\phi + \psi$, the sum of the metric potentials. Through the modified growth of structure and geometry via the lensing integral cosmic shear can constrain both
the ratio of metric potentials and modifications to the Poisson equation.
In contrast probes such as galaxy redshift surveys and galaxy peculiar velocities depend only on the Newtonian potential, $\psi$.
WL constraints also have the benefit that they probe the dark matter distribution directly, avoiding the impact of galaxy biasing. Joint constraints from combining multiple probes can help break degeneracies between
free parameters, producing the tightest constraints.

Cosmic shear is the name given to weak gravitational lensing in random patches of the sky and was first detected a decade ago \citep{Kaiser:2000if,Wittman:2000tc,van_Waerbeke:2000rm,Bacon:2000yp}.
The latest constraints on cosmology come from the Hubble Space Telescope COSMOS survey \citep{schrabbackea09,massey_growth2007_mnras}, and the Canada-France-Hawaii Telescope Legacy Survey (CFHTLS) \citep{fuea08_mnras}. In addition the 100 square degree survey \citep{benjaminea07} combines data from several smaller surveys \citep{hoekstraea2002RCS, hetterscheidt2007, lefevreea04, hoekstraea2002RCS}.

Cosmic shear has been identified as the method with the most potential to uncover the nature of dark energy \citep{detf,Peacock:2006kj} and therefore a number of surveys are planned with a major cosmic shear component. Upcoming ``Stage III" projects include Kilo-degree Survey (KIDS) on the Very Large Telescope (VLT) Survey Telescope (VST), the Panoramic Survey Telescope and Rapid Response System (Pan-STARRS) project, the Subaru Meaurement of Images and Redshifts (HSC) survey using HyperSUPRIMECam, the Dark Energy Survey (DES) on the Blanco Telescope. More ambitious ``Stage IV"
imaging projects are the Large Synoptic Survey Telescope (LSST) ground-based project and in space the proposed European Space Agency mission Euclid and the NASA proposed Wide-Field Infrared Survey Telescope (WFIRST).

Greater accuracy demands better treatment of systematic effects. One of the main systematics in cosmic shear studies is the contamination by the intrinsic alignment
(IA) of galaxy shapes. This systematic manifests in two main ways, in the first physically close galaxies have preferentially aligned ellipticities, this is known as the Intrinsic-Intrinsic (II) correlation. In the second galaxies close on the sky by separated radially are anti-correlated as the foreground gravitational potential shapes the closer galaxy while also gravitationally lensing the background galaxy, this is called the Gravitational-Intrinsic (GI) correlation. The former adds to the cosmic shear signal we wish to measure, the latter subtracts from it.
 In an earlier paper \citep{MGPaper1}, herein ``LBKB", we established how, for a fixed survey specification, the inclusion of a realistic IA model and the inherent uncertainties in the model significantly degrade the constraining power of a cosmic shear survey for dark energy and modified gravity cosmological models. This work deepened the conclusions of \citet{bernstein_2008,joachimi_bridle_2009} and expanded them into the MG domain.

What is encouraging, however, is that the use of position-position correlations, $nn$, and particularly position-shear, $n\epsilon$, cross-correlations can go some way towards mitigating the impact of IAs. This data is already collected by a standard WL survey. In the presence of IAs, constraints from $\epsilon\epsilon + n\epsilon + nn$ are roughly twice as strong as those from $\epsilon\epsilon$ alone. Until now, the discussion of optimal survey parameters has generally assumed that $\epsilon\epsilon$ correlations alone will be included, without IAs. In this paper we investigate the impact of survey strategy and design on different combinations of probes, with and without IAs. This is a continuation of the cosmic shear survey optimisation work begun by, among others, \citet{amarar07,bridleandking,joachimi_bridle_2009,mahh06,Huterer:2005ez,huterertbj06}

In this paper we expand on the work on LBKB to fully understand how the observing strategy of a given survey is central to the type of data and the quality of results that survey will produce. Different survey geometries with the same instrument will produce different statistical errors due to a different balance between survey characteristics, the most important for cosmic shear being survey area galaxy number density on the sky
and median redshift of the galaxy distribution.
Other properties such as required redshift accuracy are important guidelines for instrument designers and those preparing analysis pipelines and follow-up or calibration studies \citep{amarar07}.  While more area and increased number density will always be desirable, it is important to keep firm goals in mind when producing desiderata for future surveys which will always be limited by technology and finite observing time. In particular learning about dark energy and modified gravity may benefit from different survey strategies and call for the prioritisation of different properties.

The paper is organised as follows: in section \ref{sec:setup}, we summarise our standard cosmology, fiducial surveys and models for deviations from GR, Intrinsic Alignments and bias parameterisation, as well as our figures of merit for DE and MG. In section \ref{sec:discussion} we investigate the sensitivity of dark energy and modified gravity constraints to changes in survey specification. In section \ref{sec:area} we present the effect of varying survey area and the impact of finite survey time on overall strategy and the ability to constrain MG. The importance of photometric redshift accuracy is addressed in sections \ref{sec:photo-zs} and \ref{sec:zpriors}.
 Conclusions are made in section \ref{sec:conclusions}.

\section{Cosmological Set-Up}
\label{sec:setup}

The paper deals with the impact of various aspects of survey strategy and redshift quality on the power of cosmic shear and galaxy position information to constrain dark energy and deviations from General Relativity. In ~\ref{sec:setup-cosmicshear} we summarise our cosmic shear formalism, ~\ref{sec:setup-IAs} extends it to include intrinsic alignments and ~\ref{sec:setup-galaxys} adds galaxy position auto- and cross-correlations and introduces a coherent biasing formalism for IAs and galaxy bias. ~\ref{sec:setup-MG} reviews the MG parameterisation we use and how it enters our angular power spectrum integrals.
In ~\ref{sec:setup-cosmology} we summarise some basic cosmology we use throughout the paper and ~\ref{sec:setup-surveys} describes the fiducial survey parameters we use in the following sections.

\subsection{Cosmic Shear}
\label{sec:setup-cosmicshear}
Cosmic shear is the shape-distortion induced in the image of a distant galaxy due to the bending of its light by gravity as it passes massive structure in the Universe.
If we assume General Relativity holds then we can define the cosmic shear angular power spectrum under the Limber approximation as
\begin{equation}
C_{ij}^{GG}(l) = \int \frac{d\chi}{\chi^2} W_{i}(\chi)W_{j}(\chi) P_{\delta\delta}(k,z)
\label{eqn:GG_integral}
\end{equation}
where $P_{\delta\delta}(k,z)$ is the three dimensional matter power spectrum, $\chi$ is the comoving distance in units of $h^{-1}$Mpc and $W(\chi)$ is the lensing efficiency function
\begin{equation}
W_{i}(\chi) = \frac{4\pi G}{c^2}\rho(z)a^{2}(z)\chi \int d\chi' n_{i}(\chi')\frac{\chi'-\chi}{\chi'}
\end{equation}

\subsection{Intrinsic Alignments}
\label{sec:setup-IAs}
The intrinsic alignment (IA) of galaxy ellipticities is a prime contaminant to the measured cosmic shear signal. A naive approach to cosmic shear assumes that galaxy's intrinsic ellipticities are randomly distributed on the sky so, when we average over observed ellipticity in a small patch, intrinsic ellipticity cancels and we are left with the induced shear.

Unfortunately, this assumption is invalid because galaxy ellipticities are aligned due to two effects
arising from the same physical intrinsic alignment origin. Physically close galaxies tend to align with the local gravitational tidal field and so are positively correlated with each other, this is the Intrinsic-Intrinsic (II) alignment. Background galaxies can have their light lensed by foreground gravitational fields which align the intrinsic ellipticity of foreground galaxies. This induces an anti-correlation and is the gravitational-intrinsic (GI) alignment.

We follow the procedure of LBKB and implement IAs using an updated version of the corrected LA model of \citet{hiratas04}, incorporating their correction of \emph{erratum 2010} and assuming that all IA physics occurs at a high-z epoch of galaxy formation hence the IA signal depends only on the linear matter power spectrum rather than the subsequent nonlinear evolution. The rest of the LA model enters as a factor of
\begin{equation}
b_I = -C_{1}\rho_{m}(z=0)
\label{Isource}
\end{equation}
where $\rho_m(z=0)$ is the matter density today and  $C_1 =\textrm 5\times10^{-14}(h^{2}M_{\odot}Mpc^{-3})^{-1}$ is the amplitude of the IA term, normalised to redshift zero \citep{bridleandking}. This factor appears in the IA projected angular power spectra, linearly in the Gravitational-Intrinsic (GI) correlation and squared for the Intrinsic-Intrinsic (II). The window function associated with IAs is the galaxy redshift distribution, $n(z)$, see table \ref{table:Cls} for the full angular power spectra equations.

In section \ref{sec:setup-galaxys} we expand our notation to allow nuisance parameters to parameterise our knowledge of IAs, galaxy bias and their cross-correlations. In this more general case $b_I$ given in eqn. \ref{Isource} becomes the fiducial value of a variable IA term which appears quadratically in the II power spectrum and linearly in the GI and gI power spectra.

The total observed lensing signal is then the sum of the cosmic shear and the IA terms
\begin{equation}
C_{l}^{\epsilon\epsilon} =C_{l}^{GG} + C_{l}^{II} + C_{l}^{GI}.
\end{equation}

\subsection{Galaxy Position Data}
\label{sec:setup-galaxys}

\begin{table*}
\caption{Summary of the projected angular power spectra considered in this work.}
\centering
\begin{tabular}[t]{l||l}
Correlation & 2D PS \\
\hline
\hline
Shear                                   & $C_{ij}^{GG}(l) = \int \frac{d\chi}{\chi^2}W_{i}(\chi)W_{j}(\chi)\left[ Q(z)\frac{1+R(z)}{2} \right] ^2 P_{\delta\delta}(k,z) $ \\
Intrinsic-shear                         & $C_{ij}^{GI}(l) = \int \frac{d\chi}{\chi^2}W_{i}(\chi)n_{j}(\chi)Q(z)Q(z_f)R(z_f)(1+\frac{R(z)}{2})b_{I}(k,z)r_{I}(k,z)\frac{\sqrt{P_{\delta\delta}(k,z_f)P_{\delta\delta}(k,z)}}{D(z_f)}$\\
Intrinsic                               & $C_{ij}^{II}(l) = \int \frac{d\chi}{\chi^2}n_{i}(\chi)n_{j}(\chi)b_{I}^{2}(k,z)Q^{2}(z_f)R^{2}(z_f) P_{\delta\delta}(k,z) $\\
Galaxy clustering                       & $C_{ij}^{gg}(l) = \int \frac{d\chi}{\chi^2}n_{i}(\chi)n_{j}(\chi)b_{g}^{2}(k,z) P_{\delta\delta}(k,z)$\\
Clustering-shear                        & $C_{ij}^{gG}(l) = \int \frac{d\chi}{\chi^2}n_{i}(\chi)W_{j}(\chi) Q(z)\frac{1+R(z)}{2}b_{g}(k,z)r_{g}(k,z) P_{\delta\delta}(k,z)$\\
Clustering-intrinsic                    & $C_{ij}^{gI}(l) = \int\frac{d\chi}{\chi^2}n_{i}(\chi)n_{j}(\chi)Q(z_f)R(z_f)b_{g}(k,z)b_{I}(k,z)r_{g}(k,z)r_{I}(k,z) \frac{\sqrt{P_{\delta\delta}(k,z_f)P_{\delta\delta}(k,z)}}{D(z_f)}$\\
Galaxy ellipticity (observable)         & $C_{ij}^{\epsilon \epsilon} =  C_{ij}^{GG}+ C_{ij}^{II}+ C_{ij}^{GI}$  \\
Galaxy number density (observable)      & $C_{ij}^{n n} = C_{ij}^{gg}$          \\
Number density-ellipticity (observable) & $C_{ij}^{n \epsilon} =  C_{ij}^{gI} +  C_{ij}^{gG}$          \\
\end{tabular}
\label{table:Cls}
\end{table*}

A cosmic shear survey contains galaxy position information (angular position on the sky and redshift) as well as measurements of galaxy shear. \citet{joachimi_bridle_2009} provide a formalism for including this additional information in cosmological parameter constraints and show how this extra information can serve to partially mitigate the impact of IAs. This approach follows from the work of \citet{zhang08,hu_jain_2004}

The extra observables we use are galaxy position-position power spectra and the position-shear cross-spectra, defined analogously to the shear-shear power spectrum:
\begin{align}
C_{l}^{nn} &= C_{l}^{gg} \\
C_{l}^{n\epsilon} &= C_{l}^{gI} + C_{l}^{gG}.
\end{align}
We now assign nuisance parameters to each component $C_l$ in a self-consistent way as explained in LBKB. The four bias
function are $b_g$, $b_I$, $r_g$ and $r_I$. $b_g$ and $b_I$ model bias from galaxy position and IAs amplitudes, and their cross-correlations are modelled by $r_g$ and $r_I$. This unified approach was first introduced by \citet{bernstein_2008}.
Table \ref{table:Cls} summarises the full power spectra equations, consistent with the definitions in LBKB eqns. 37, 38, 43 and 44.

We let each bias parameter vary in amplitude and as a function of scale
and redshift using a $N_k \times N_z$
grid of free parameters interpolated over
$k$-,$z$-space, i.e. each nuisance factor, $b_X = A_x Q_{X}(k,z)$ is the product of a variable constant amplitude parameter, $A_X$, and a variable grid, $Q_{X}(k,z)$ in $k$ and $z$.  Throughout this paper the grid size is set to the fiducial value of $N_k = N_z = 5$, which means marginalisation over 104 nuisance parameters when the full $\epsilon\epsilon+n\epsilon+nn$ probe combination is considered in the presence of IAs.
The nuisance parameters multiply linearly into the angular power spectra integrands and each power spectrum depends on a subset of the bias parameters as follows:
\begin{equation}
\begin{array}{cccc}
C_{l}^{GG}: & -\phantom{a} ,& C_{l}^{gG}: & b_{g}r_{g} \\
C_{l}^{II}: & b_{I}b_{I}, & C_{l}^{gI}: & b_{g}b_{I}r_{g}r_{I} \\
C_{l}^{GI}: & b_{I}r_{I}, & C_{l}^{gg}: & b_{g}b_{g}
\end{array}
\end{equation}
As mentioned in the previous section, the fiducial value of $b_I$ is given by eqn. \ref{Isource}. $b_g$ has fiducial value 1 while $r_g$ and $r_I$ vary around the fiducial value 0.9 to avoid more than perfect cross-correlation.

We have ignored the effect of lensing magnification and followed LBKB and \citet{joachimi_bridle_2009} in applying a cut on multipole $l$ to any redshift bin combination $ij$ which includes galaxy position information (i.e. $nn$ or $n\epsilon$) according to $\ell_{max}(i)=k_{lin}^{max}(z_{med}^{(i)})\chi(z_{med}^{(i)})$. This aims to account for uncertainties in the galaxy bias model at small scales.
For $n\epsilon$ bin pairs, the cut is made on the galaxy, i.e. $n$, bin. For position-position, $nn$, pairs there is a choice of bin on which to apply the cut. We follow \citet{joachimi_bridle_2009} in making the optimistic choice and cutting on the higher redshift bin. Shear-shear, $\epsilon\epsilon$, pairs do not depend on galaxy bias and are therefore used up to the full range in $l$.

\subsection{Modified Gravity}
\label{sec:setup-MG}

There are a large number of modifications or extensions of Einstein's general theory of relativity
on cosmic scales which come under the general heading of Modified Gravity theories. These can be motivated by the presence of extra dimensions, as in DGP, or extra degrees of freedom compared to the GR action equation, as in f(R), TeVeS etc. \citep{DGP, carrollea_fR, skordis_teves, jain_khoury}. As in LBKB, we do not assume a particular modified theory of gravity but rather concentrate on ``trigger parameters'' whose deviation from their GR values would indicate the presence of some physics beyond that in the standard GR picture.

In the conformal newtonian gauge the metric for a flat FRW spacetime is written
\begin{equation}
ds^2 = -a(\tau)^2 \left[ 1 + 2\psi(x,t)\right]d\tau^2 + a(\tau)^2\left[1-2\phi(x,t) \right]dx^2
\end{equation}
where $\psi$ and $\phi$ are the scalar potentials which describe perturbations to the time- and space-parts of the metric respectively. We parameterise deviations from GR through two parameters, $Q$ and $R$. One alters the way the Newtonian potential responds to mass via the Poisson equation,
\begin{equation}
k^2\psi(x,t) = -4\pi GQ \rho a^2 \delta
\end{equation}
and the other modifies the ratio of the metric potentials
\begin{equation}
\psi(x,t) = R\phi(x,t)
\end{equation}
$Q$ and $R$  are assumed to be scale-independent and vary with redshift as
\begin{align}
Q &= (Q_0-1) a^s \\
R &= (R_0-1) a^s
\end{align}
where a is the scale-factor and $Q_0$,$R_0$ are the free parameters we vary for MG, and $s=3$. We are interested in MG theories which explain the acceleratin expansion of the Universe observed at late times. $s=3$ allows any modification to ``turn-on'' and late times and avoids violating early Universe constraints from the Cosmic Microwave Background (CMB) and Big Bang Nucleosynthesis (BBN).

The modified gravity parameters enter the projected angular power spectra in different combinations, as shown in Table \ref{table:Cls}, due to their different dependencies on the mater density field. As well as modifying the angular power spectra, Q and R enter the angular power spectrum integrals via their response to the metric potentials through the linear growth function. The potential for weak lensing to constrain deviations from GR has previously been noted in \citet{Bean:2010zq,Laszlo:2007td,Beynon:2009yd}.

The full projected angular power spectra, including IAs, MG and bias parameters, is summarised in Table \ref{table:Cls}.
Note also that deviations from GR will also manifest in the growth function which produces the matter power spectra in Table \ref{table:Cls}. We include modifications to growth via a ratio of MG/GR power spectra calculated using
the modified version of CAMB used for LBKB.
For ease, LBKB also provide a fitting function to compute this ratio over a range of $Q_0$, $R_0$ and $s$ values.

\begin{table}
\begin{center}
\begin{tabular}{|l|l|l|}
\hline
\multicolumn{3}{|c|}{Fiducial Survey Paramters} \\
\hline
Parameter & Stage III & Stage IV \\
\hline
Area            & 5,000deg$^2$ & 20,000deg$^2$ \\
$n_g$           & 10            & 35 \\
$\sigma_\gamma$ & 0.23          & 0.35 \\
$N_z$           & 5             & 10 \\
$\delta_z$      & 0.07          & 0.05 \\
$f_{cat}$       & 0             & 0 \\
$\Delta_z$      & 1             & 1 \\
$z_0$           & 0.8/1.412     & 0.9/1.412 \\
$\alpha$        & 2             & 2 \\
$\beta$         & 1.5           & 1.5 \\
\hline
\end{tabular}
\caption{Fiducial survey parameter values for the DES-like (stage III) and Euclid-like (stage IV) surveys.}
\label{fig:DESEuclid_table}
\end{center}
\end{table}

\subsection{General Cosmology}
\label{sec:setup-cosmology}

Throughout this paper we assume a flat $\Lambda$CDM cosmological model with fiducial parameter values equal to the WMAP7 best-fit values:
$\Omega_m = 0.262$, $\Omega_b = 0.0445$, $w_0 = -1$, $w_a = 0$, $\sigma_8 = 0.802$, $h = 0.714$, $n_s =0.969$,  $\Omega_\nu = 0$.
Where $\Omega_m$,$\Omega_b$ and $\Omega_\nu$ are the dimensionless matter, baryon and neutrino densities respectively, $w_0$ and $w_a$ are the dark energy equation of state parameters \citep{detf},$\sigma_8$ is the normalisation of the linear matter power spectrum, $h$ is the dimensionless Hubble parameter today and $n_s$ is the power law of the primordial power spectrum.
The linear matter power spectrum is given by the fitting formula of \citet{eisensteinhu97}, with nonlinear corrections  from \citet{smithea03}.

When we treat MG we assume that the background expansion of the Universe is consistent with our fiducial
dark energy model and deviations due to non-GR physics enter through the growth of structure from initial perturbations. Unless otherwise specified all plots include CMB priors from a Planck-type survey
as described in LBKB. We divide our survey galaxy redshift distributions into redshift slices with equal galaxy number density to allow redshift tomography where auto- and cross-correlations of these redshift bins are considered, allowing us to measure redshift evolution of the observables.

We calculate constraints on cosmological parameters using the Fisher Matrix formalism
\begin{equation}
F_{\mu\nu} = \sum_{m,n}^{N_d} \sum_{l}^{N_{l}^{max}}\frac{\partial\mathscr{D}_{m}(l)}{\partial p_{\mu}}\textrm{Cov}_{mn}^{-1}(l)\frac{\partial\mathscr{D}_{n}(l)}{\partial p_{\nu}}
\end{equation}
where $m$, $n$ label tomographic redshift bins,  $\mathscr{D}_{m}(l)$ is the data vector, in our case some combination of the angular power spectra $C_{l}^{\epsilon\epsilon}$, $C_{l}^{n\epsilon}$ and $C_{l}^{nn}$. $\textrm{Cov}_{mn}(l)$ is the covariance matrix, defined as in \citet{joachimi_bridle_2009}, $p_{\mu}$ are the parameters varied and $N_d$ is the number of independent combinations of tomographic bins. Unless otherwise stated the Fisher matrix varies the cosmological  parameters: $p = \{\Omega_m,w_0,w_a,h,\sigma_8,\Omega_b,n_s\}$.

Using the Fisher matrix formalism the lower limit on the minimum variance bound on the error of a parameter $p_\mu$
marginalised over all other parameters of interest, is given by $\sigma(p_\mu) = \sqrt{(F^{-1})_{\mu\mu}}$.
We quote results in terms of the DETF FoM for Dark Energy
\begin{equation}
FoM_{DE} = \frac{1}{\sqrt{\textrm{det}\left[(F^{-1}_{GR})_{w_{0},w_{a}}\right]}}.\label{DEFoM}
\end{equation}
Here $(F^{-1}_{GR})_{w_{0},w_{a}}$
is the $2\times 2$ submatrix of the inverted Fisher matrix for cosmological and nuisance parameters excluding the MG parameters are excluded (these are fixed at their GR values).  The FoM is proportional to the inverse of the area of the constraint contour in $w_0 - w_a$ space.

By analogy we define a MG FoM as
\begin{equation}
FoM_{MG} = \frac{1}{\sqrt{\textrm{det}\left[(F^{-1})_{Q_{0},Q_{0}(1+R_{0}/2)}\right]}}.
\label{MGFoM}
\end{equation}
where we have inverted the full fisher matrix including the modified gravity parameters
and dark energy parameters.

\subsection{General Survey Parameters}
\label{sec:setup-surveys}
LBKB investigated the degeneracy between IAs and MG parameters for a fixed survey specification. It was found that the inclusion of a realistic IA model reduced the FoM (dark energy or MG) for a typical stage-IV cosmic shear survey by $\sim$70\%. The effect could be substantially mitigated by the inclusion of galaxy position data and galaxy-shear cross-correlations. The constraining power of $\epsilon\epsilon+n\epsilon+nn$, including IAs, is roughly double that of $\epsilon\epsilon$, including IAs, alone. We showed that our results were robust to different numbers of free nuisance parameters, accounting for flexibility in the galaxy clustering and IA bias models. The results were broadly similar for attempts to constrain dark energy equation of state parameters and deviations from GR.

In this paper we vary certain survey parameters around a fiducial cosmic shear survey set-up, corresponding to
Stage IV project. This survey has an area = 20000deg$^2$,  number density of galaxies projected onto the sky, $n_g = 35$ arcmin$^{-2}$, and galaxy intrinsic ellipticity dispersion, $\sigma_\gamma = 0.35$, a Gaussian photometric redshift scatter of width $\delta_z = 0.05$ which is related to the (redshift dependent) rms photometric dispersion via $\sigma_z(z) = (1+z)\delta_z$, with a fraction of catastrophic outliers, $f_{\textrm{cat}} = 0$. We analyse the results using cosmic shear tomography with 10 tomographic redshift bins of equal number density.

The redshift distribution of galaxes, $n(z)$ is
assumed to be given by a Smail-type distribution \citep{smailef94}
\begin{equation}
\label{eq:smail}
n(z) \propto z^\alpha exp \left( -\frac{z}{z_0} \right)^{\beta}
\end{equation}
with $\alpha = 2$, $\beta = 1.5$ and $z_0 = z_{m}/1.412$, where the median redshift, $z_m$ = 0.9.

For comparison purposes, some results are also presented for the stage-III Dark Energy Survey (DES), due to see first light in early 2012. Table \ref{fig:DESEuclid_table}  summarises the specifications for both surveys.

\begin{figure*}
  \begin{flushleft}
    \centering
       \includegraphics[width=4.5in,height=4.5in]{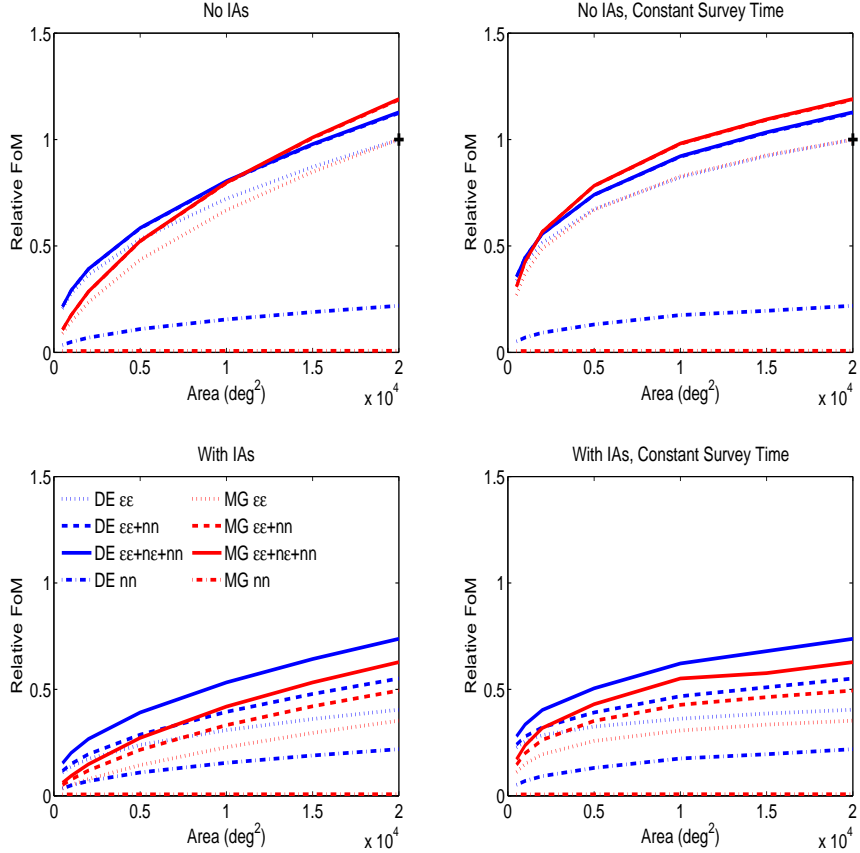}
\caption{Figures showing the variation with survey area of a relative Stage IV survey FoM for dark energy (DE) parameters $w_0,w_a$ [blue
 lines], and modified gravity (MG) parameters $Q_0, Q_0(1+R_0)/2$ [red lines], as described in (\ref{DEFoM}) and (\ref{MGFoM}), relative to a `baseline' FoM.  The baseline includes shear  auto-correlations `GG' alone, excluding IAs, over a 20,000 square degree survey. All FoMs contain priors from a Planck-like CMB survey.
 The figures show  relative FoMs for an optimistic scenario, in which uncertainities in the IA model have been excluded, [top panels] and a conservative scenario, in which uncertainties in the IA model are marginalized over using a $N_k=N_z=5$ gridded bias model [lower panels]. The baseline model (relative FoM=1) is shown as a black cross in the top panels. We consider the impact on the relative FoM of increasing survey area, by increasing survey time for fixed limiting magnitude [left panels] and fixing the survey time to trade-off survey area and depth [right panels].  Four data combinations are considered in each panel, shear-shear correlations `$\epsilon\epsilon$' [dotted lines], galaxy-galaxy positions `$nn$' alone [dot-dash], '$\epsilon\epsilon$+$nn$' [dashed] and when shear-position cross-correlations `$n\epsilon$' are included [full lines]. For the galaxy position correlations a $N_k=N_z=5$ gridded bias model is used throughout. }
\label{fig:varyA}
  \end{flushleft}
\end{figure*}

 \begin{figure*}
 \center
       \includegraphics[width=4.5in,height=4.5in]{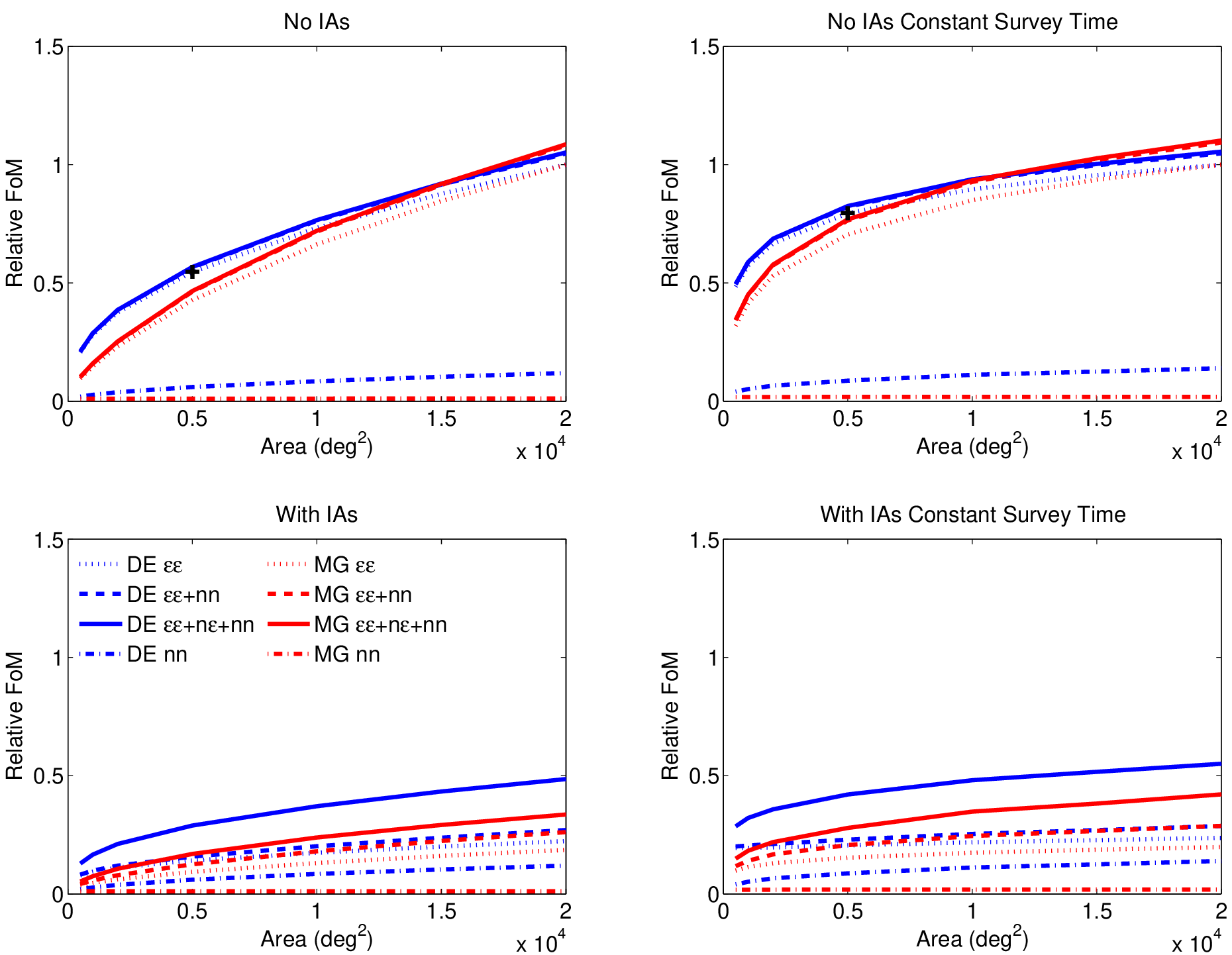}
 \caption{
 As in figure \ref{fig:varyA} but for a Stage-III large scale structure survey with fiducial survey area of 5000 square degrees.}
 \label{fig:Aconstt_DES}
 \end{figure*}

\section{Survey specifications and forecast constraints}
\label{sec:discussion}
In this section we analyse the sensitivity of the prospective dark energy and modified gravity figures of merit on variations in survey specification, principally focusing on their relationship to the survey area and depth, and the precision of the photometric redshift measurements when uncertainties in how intrinsic alignments are included in the modeling.

Throughout, we  quantify constraints using the dark energy (DE) and modified gravity (MG) figures of merit in (\ref{DEFoM}) and (\ref{MGFoM}) normalised relative to a `baseline' figure of merit for the shear-only auto-correlations `$GG$', with IAs excluded, using the fiducial survey described in Table \ref{fig:DESEuclid_table}.

\subsection{Survey area}
\label{sec:area}

Figs.~\ref{fig:varyA} and \ref{fig:Aconstt_DES} present the key results showing how the constraining power of a survey, for dark energy equation of state and modified gravity parameters, varies with the survey area. We consider two alternative approaches, firstly a simple option where the increased area derives from an increase in survey time, and secondly a more realistic option when survey time is fixed, and increased survey area is achieved by a reduction in survey depth/ limiting flux.

Let us first consider the case where all survey parameters are kept fixed except for the survey area. We focus on the quantitative implications for the Stage IV survey, the results for a Stage III survey, presented in figure \ref{fig:Aconstt_DES} are qualitatively similar.

When IAs are excluded from the analysis the improvement in FoM with increasing area can be roughly described by a power law dependence FoM$ \propto$Area$^x$ with $x=0.4$ and 0.7  for CMB+shear correlations alone for the DE and MG FoMs. The MG FoM is slightly more sensitive to changes in area than the DE FoM. The DE index is lower than  that  reported in \citet{amarar07} only because of the inclusion of, survey area independent, CMB priors.

As discussed in LBKB, including IAs in the analysis deteriorates both the DE and MG FoMs, with the MG FoM more negatively impacted. For the fiducial Stage IV survey the FoMs are reduced by 60 and 65 \% relative to when IAs are excluded, consistent with LBKB. When IAs and their associated uncertainties  are marginalized over, the power law scaling is weaker, with $x=0.35$ for DE, so that the relative degradation of the FoM becomes more pronounced as survey area increases. This reflects that the IAs are astrophysical systematics that are not significantly removed by increasing the survey area.

Because of the uncertainties in the galaxy bias, through the gridded bias and cross-correlation coefficients, $b_g$ and $r_g$, the cosmological information from the galaxy position data is massively suppressed. For the fiducial full survey area FoMs for CMB+nn are 22 and 1\% of the CMB+pure lensing signal, for the scenarios with and without IAs respectively.

 If one adds in galaxy position autocorrelations to shear autocorrelations, `$nn+\epsilon\epsilon$', we find a 12 and 36 \%  improvement in the DE and with and without IAs respectively, and 19 and 40\% for MG FoMs. While the cross-correlations, `$nn+n\epsilon+\epsilon\epsilon$', give only a small improvement when IAs are excluded, they provide an 34\% improvement on `$nn+\epsilon\epsilon$' when uncertainties in the IAs are included. This results from the $n\epsilon$ correlations breaking the degeneracies between the cosmological and IA amplitude nuisance parameters.

In reality most surveys have a fixed observing time.
This inevitably leads to a trade off between total survey area and survey depth, which impacts both number density of lensed galaxies, $n_g$, and median redshift achieved, $z_m$.
To implement the fixed survey time option, we use a fitting formula for the number of galaxies per unit redshift per square arcminute
\begin{equation}
n(z) = \Sigma_0 \times \frac{3 z^2}{2 z_0^3} \exp{ \left[ -
\left( \frac{z}{z_0} \right)^{3/2} \right] }
\label{eqdndz}
\end{equation}
(e.g. \citet{baugh_efstathiou})
and we interpolate the numbers from Table 1 of \citet{BlakeBridle}
using
\begin{align}
z_0 & = 0.055 (r_{\rm lim}-24)+0.39 \nonumber
\\
\Sigma_0 &=  \frac{35400}{60^2} \left(\frac{r_{lim}}{24}\right)^{19},
\label{rlimdef}
\end{align}
where $r_{\rm lim}$ is the limiting apparent magnitude in the SDSS $r$ filter and $\Sigma_0$ is an overall surface density in units of deg$^{-2}$. We use (\ref{rlimdef}) to scale the number density of galaxies with area relative to the fiducial values.
We estimate the change in the limiting magnitude, $\Delta_{\rm mag}$, with area in the usual way as
\begin{equation}
\Delta_{\rm mag} = -1.25log_{10}\left(\frac{A}{A_{\rm fid}}\right) .
\end{equation}

As one might expect, the benefits of going to larger survey area are now less strong because of the lost depth, with the shear-only power law dependence weakening to $x$ = 0.3 for DE and 0.35 for MG.
The
baseline 
result for cosmic shear analyses is that it is still better to go to larger survey area despite the lost depth.
Figure 4 of \citet{amarar07} suggests that, in the absence of a
CMB prior and IA modeling uncertainties, the FoM is roughly proportional to the square root of the area for fixed observing time. We find that  with the inclusion of CMB priors,
we obtain a similar result although the dependence on area is slightly weakened as in the constant depth case. When the galaxy position information is added the same story holds, but the dependence on area is slightly stronger.

When intrinsic alignments are added the improvement with increased area is significantly reduced, particularly in the case of $\epsilon\epsilon$-only which reduces to a power law of $x\sim$0.15 for DE.
A survey with an area of $5,000$deg$^2$ gathers $\sim 80\%$ of the DE information of a $\sim 20,000$deg$^2$ survey. This is because intrinsic alignments dominate over cosmic shear at very low redshift, and by marginalising out the intrinsic alignment nuisance parameters we effectively remove the information at low redshift. Therefore we expect any cosmological figure of merit to eventually reduce in the (unphysical) limit of very large area and small depth.

The galaxy clustering alone constraints are insensitive to the intrinsic alignments and therefore increase as usual.
When using the full $\epsilon\epsilon+n\epsilon+nn$ dataset there is more benefit to increased area even with the tradeoff of shallower depth, although in this case a 5000 square degree survey still gathers $\sim 70\%$ of the information
in a $\sim 20,000$deg$^2$ survey. Note that the exact details of these conclusions will depend on the fiducial intrinsic alignment model and the number of intrinsic alignment and bias nuisance parameters marginalised over.
Without galaxy position-shear cross correlations the constraints bear out the same trend in relative FoM with area, just with an overall lower magnitude.

 We repeated the exercise using the default survey parameters for a stage III survey in Fig.~\ref{fig:Aconstt_DES}.
 We vary the survey area for fixed survey time, and cover the same range as in Figure \ref{fig:varyA}.
 The survey depth for each survey time is correspondingly shallower, so that we have 10 galaxies per square arcminute for the 5000 square degree survey area.
The left-hand plots in Fig.~\ref{fig:Aconstt_DES} tell a similar story to those in Fig.~\ref{fig:varyA}. When a constant survey time is considered there is a more pronounced tailing off of improvement with area above $\sim 5-10,000$ deg$^2$ than was the case for a stage IV survey. As the area is increased above 5000deg$^2$ the number density of galaxies falls from its fiducial value of 10 arcmin$^{-2}$. For larger areas the number density becomes very low and shot noise rapidly increases at small scales. The loss of information from non-linear scales has a strong impact on the constraining power of the survey, limiting the improvement with increasing area. The fact that $nn$ alone does not display this behaviour, displaying the same behaviour for fixed depth or survey time, supports our analysis- the galaxy position information is already subject to a stringent cut on $l$, removing the nonlinear information and protecting nn from the impact of shot noise.

\begin{figure*}
  \begin{flushleft}
    \centering
       \includegraphics[width=4in,height=4.35in]{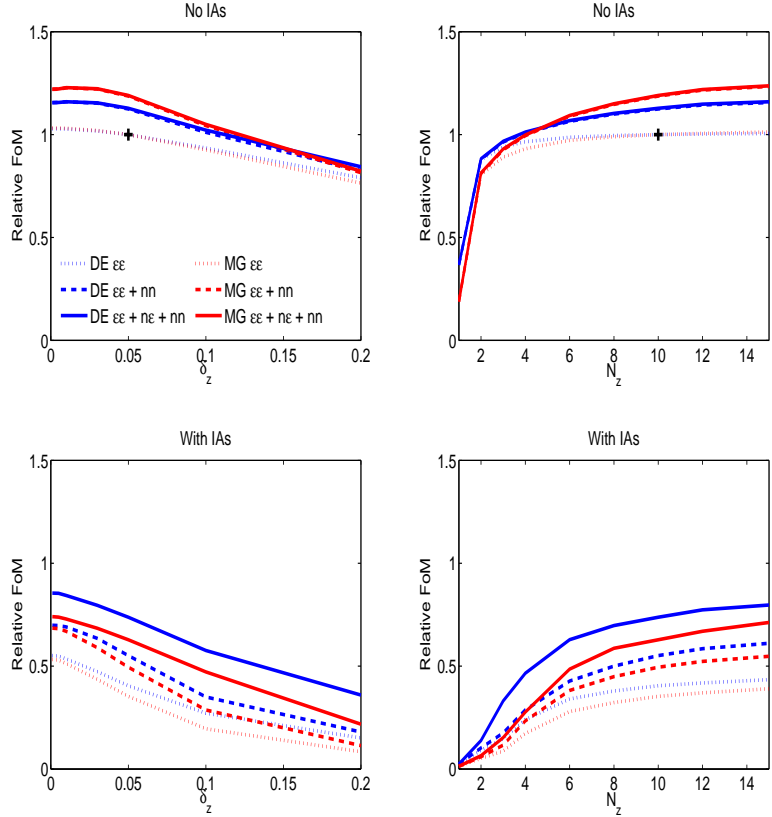}
\caption{Figures showing the variation with photometric redshift error ($\delta_z$) [left panels] and number of tomographic bins ($N_z$) [right panels] of a Stage IV survey FoM for dark energy (DE) parameters $w_0,w_a$ [blue lines], and modified gravity (MG) parameters $Q_0, Q_0(1+R_0)/2$ [red lines], as described in (\ref{DEFoM}) and (\ref{MGFoM}), relative to a `baseline' FoM [black cross].
 All FoMs contain priors from a Planck-like CMB survey. The figures show  relative FoMs for an optimistic scenario, in which uncertainities in the IA model have been excluded, [top panels] and a conservative scenario, in which uncertainties in the IA model are marginalized over using a $N_k=N_z=5$ gridded bias model [lower panels].  Three data combinations are considered in each panel, shear-shear correlations `$\epsilon\epsilon$' [dotted lines], '$\epsilon\epsilon$+$nn$' [dashed] and when shear-position cross-correlations `$n\epsilon$' are included [full lines]. For the galaxy position correlations a $N_k=N_z=5$ gridded bias model is used throughout.
 }
\label{fig:FoM_Deltaz_Nz}
  \end{flushleft}
\end{figure*}

\subsection{Photometric Redshifts}
\label{sec:photo-zs}

Accurate redshift information is essential for weak lensing tomography. As survey size increases
the number of redshifts becomes of order hundreds of millions to billions.
The expense of acquiring spectroscopic redshifts for this number of galaxies is prohibitive. As a result the next generations of WL surveys will rely on photometric redshifts.

In this paper we follow the photometric redshift model of \citet{amarar07}, including the statistical dispersion of measured photometric redshifts and catastrophic failures in redshift measurement.
The statistical dispersion is modelled assuming that measured photometric redshifts, $z_{phot}$ are described by a Gaussian probability distribtion centred on the true redshift, $z_t$.
The width of the Gaussian is
the photometric redshift error which we assume scales as $(1+z)$ and is controlled
by a paramter $\delta_z$,
\begin{equation}
P_{z_{phot}}(z) = \frac{1}{\sqrt{(2 \pi) \delta_z (1+z)}}
\end{equation}

In addition to this uncertainty in the photometric PDF galaxy redshifts can be entirely misidentified if, for example, the wavelength range examined is insufficient to identify important features in the spectrum. A misidentified redshift can be said to be assigned a redshift offset by an amount $\Delta_z$ from the true redshift. Affected galaxies are known as catastrophic outliers.

We construct a PDF to describe these catastrophic outliers, $P_{z_{cat}}$, and combine it with the scatter PDF, writing the full probability distribution
\begin{equation}
P(z_{phot}|z_t) = (1-f_{cat})P_{z_{phot}}(z_t) + (f_{cat})P_{z_{cat}}(z_t)
\end{equation}
where $f_{cat}$ is the fraction of our measured galaxies which suffer catastrophic redshift estimation failures.

The distribution of $P_{z_{cat}}$ is bimodal, reflecting the fact that galaxies can either be misidentified as being higher or lower than their true redshifts,
\begin{equation}
<P_{z_{cat}^{\pm}}> = z_{cat}^{\pm} = z_t \pm \Delta_z
\end{equation}
where $P_{z_{cat}} = P_{z_{cat}^{-}} + P_{z_{cat}^{+}}$.
The value of $\Delta_z$ will depend on the survey filters, photometric reconstruction technique and spectral properites of the survey galaxies. We assume a fiducial value of $\Delta_z = 1$ and that the uncertainty on the PDF of the population of catastrophic outliers is the same as for the slightly scattered galaxies
\begin{equation}
\sigma(P_{z_{cat}^{\pm}}) = \delta_z(1+z_{cat}^{\pm}).
\end{equation}

Fig.~\ref{fig:FoM_Deltaz_Nz} shows the variation in FoMs with $\delta_z$ and $N_z$, the number of tomographic redshift bins, for different combinations of observables, with and without IAs.
There is a parallel between varying $\delta_z$ and $N_z$, lower redshift error focuses the redshift distribution per bin into a tighter shape, allowing us to learn from changes with redshift over shorter intervals. Somewhat equivalently, increasing the number of tomographic bins gives us, in principle, the opportunity to detect changes with redshift at higher resolution.

Increasing $\delta_z$ decreases the constraining power of a survey because the loss of redshift precision ``smears out'' the redshift distribution of our sample. In particular galaxies leak between tomographic redshift bins meaning we can extract less information from tomography.
         The impact of increased $\delta_z$ on constraining DE is relatively weak. For $\epsilon\epsilon$ alone the
 DE FoM drops by only 1/4 from $\delta_z = 0$ (spectroscopic, i.e. perfect redshift precision) to $\delta_z = 0.2$, already a much higher value than the fiducial redshift uncertainties of
typical Stage-III ($\delta_z = 0.07$) and Stage-IV ($\delta_z = 0.05$) experiments. MG FoMs display very similar behaviour.

The reasons for this behaviour are relatively well understood. The lensing integral (eqn.~\ref{eqn:GG_integral}) modulates the 3D matter power spectrum by the lensing weight functions $W_{i}(\chi)$ of the redshift bins being correlated. These lensing weight functions are broad in redshift and act as kernels
smoothing the redshift information. This effect limits the usefulness of increasingly accurate redshift information and explains why the trend with $\delta_z$ is relatively shallow.

The inclusion of galaxy position information produces slightly steeper trends with increased $\delta_z$ because their angular power spectra integrals are modulated by the redshift distribution $n_{i}(\chi)$ of the bin(s) in question. This is a narrower function than the lensing weight function, providing more scope for improvement in accuracy with better redshift information.

The slightly steeper curves for the MG FoMs may come from the larger number of parameters that need to be constrained in this model, which can be done with new types of information such as increased redshift resolution. Broadly the results for modified gravity are very similar to those for dark energy.

All of the
data combinations with varying number of tomographic redshift bins, $N_z$, show the characteristic plateaux behaviour
as $N_z$ increases.
Without IAs, increasing the number of tomographic bins becomes much less effective after $N_z \sim2-3$. After this number $\epsilon\epsilon$ is effectively flat, while $\epsilon\epsilon + nn$ and $\epsilon\epsilon + n\epsilon + nn$ flatten out slowly, gaining little benefit after $N_z \sim 10-12$. This behaviour is expected as the lensing kernel and photo-z scatter smear-out redshift information and limit the ability of ever-finer tomography to improvement our knowledge of redshift behaviour.

The inclusion of IA terms for the same probes
shows the expected decrease in constraining power as we have to marginalise over IA bias terms. As we have seen previously, the MG FoMs take more of a hit due to the IA effects than their DE counterparts.

The decrease in FoM with increasing $\delta_z$ is significantly more pronounced once IAs are included, with $\epsilon\epsilon$ including IAs losing ~$67\%$ of its constraining power for DE  over the interval probed. Qualitatively we might hope to understand this from the fact that the cosmic shear and each IA term, II and GI, have a different redshift dependence. Improved redshift knowledge allows us to better discriminate between the cosmic shear signal and IA contamination. This produces stronger constraints on the IA nuisance parameters and consequently tighter constraints on DE and MG.

Similarly, the presence of IAs for increasing $N_z$ makes the plateau behaviour less pronounced and pushes the levelling off of the FoM back towards $N_z\sim 6-10$ for all probes. The use of smaller slices in redshift space gains information about the z-distribution of the IA signals and aids our constraining power. The effect is curtailed by the same fundamental limit of redshift accuracy in the projected angular power spectrum. Again the hit on FoM due to IAs is more pronounced for MG than DE but the qualitative behaviour of FoM with $N_z$ in the presence of IAs is the same for both.

\begin{figure}
  \begin{flushleft}
    \centering
       \includegraphics[width=3.5in,height=2in]{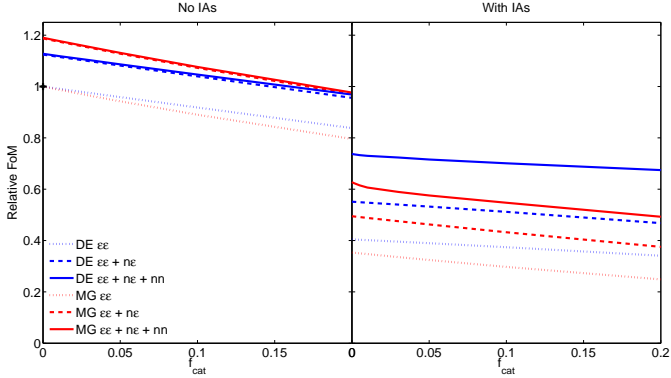}
\caption{Figures showing the variation with fraction of catastrophic redshift outliers ($f_{cat}$) of a Stage IV survey FoM for dark energy (DE) parameters $w_0,w_a$ [blue lines], and modified gravity (MG) parameters $Q_0, Q_0(1+R_0)/2$ [red lines], as described in (\ref{DEFoM}) and (\ref{MGFoM}), relative to a `baseline' FoM [black cross].
The figures show  relative FoMs for an optimistic scenario, in which uncertainities in the IA model have been excluded, [left panel] and a conservative scenario, in which uncertainties in the IA model are marginalized over using a $N_k=N_z=5$ gridded bias model [right panel].
Notation is the same as in Fig.~\ref{fig:FoM_Deltaz_Nz}}
\label{fig:FoM_fcat}
  \end{flushleft}
\end{figure}

The effect of changing the fraction of catastrophic outlier redshifts in our survey sample is shown in Fig.~\ref{fig:FoM_fcat}. As we would expect increasing the fraction of completely mis-estimated redshifts makes for poorer constraints on both MG and DE, for similar reasons to the impact of $\delta_z$.

All the lines in Fig.~\ref{fig:FoM_fcat} are relatively flat, suggesting that the projected nature of the angular power spectra and the effect of the lensing weight function continue to limit the impact of redshift information on overall constraints. No probe combination loses more than 30\% of its DE or MG FoM as $f_{cat}$ increases from zero to 0.2 which is more than twice the typical requirement value of projected stage-IV surveys.

The general pattern that MG FoMs decrease slightly more strongly than their DE equivalents with the addition of IAs is repeated here. One interesting feature is that the MG constraints for $\epsilon\epsilon$ without IAs and all the probe combinations with IAs decrease more steeply with increasing $f_{cat}$ than their DE counterparts. The difference is most pronounced for $\epsilon\epsilon$ with IAs where the MG constraints falls by twice as much over the $f_{cat}$ range as the DE FoM.

\subsection{Redshift Priors}
\label{sec:zpriors}

As well as describing the change in FoMs with the size of the photometric uncertainty, it is important to the quantify the effect of our \emph{degree of knowledge} of that uncertainty on our FoMs. In our case we follow the standard procedure of inserting this degree of knowledge as a prior on the parameter $\delta_z$.

\begin{figure}
  \begin{flushleft}
    \centering
       \includegraphics[width=3.5in,height=4in]{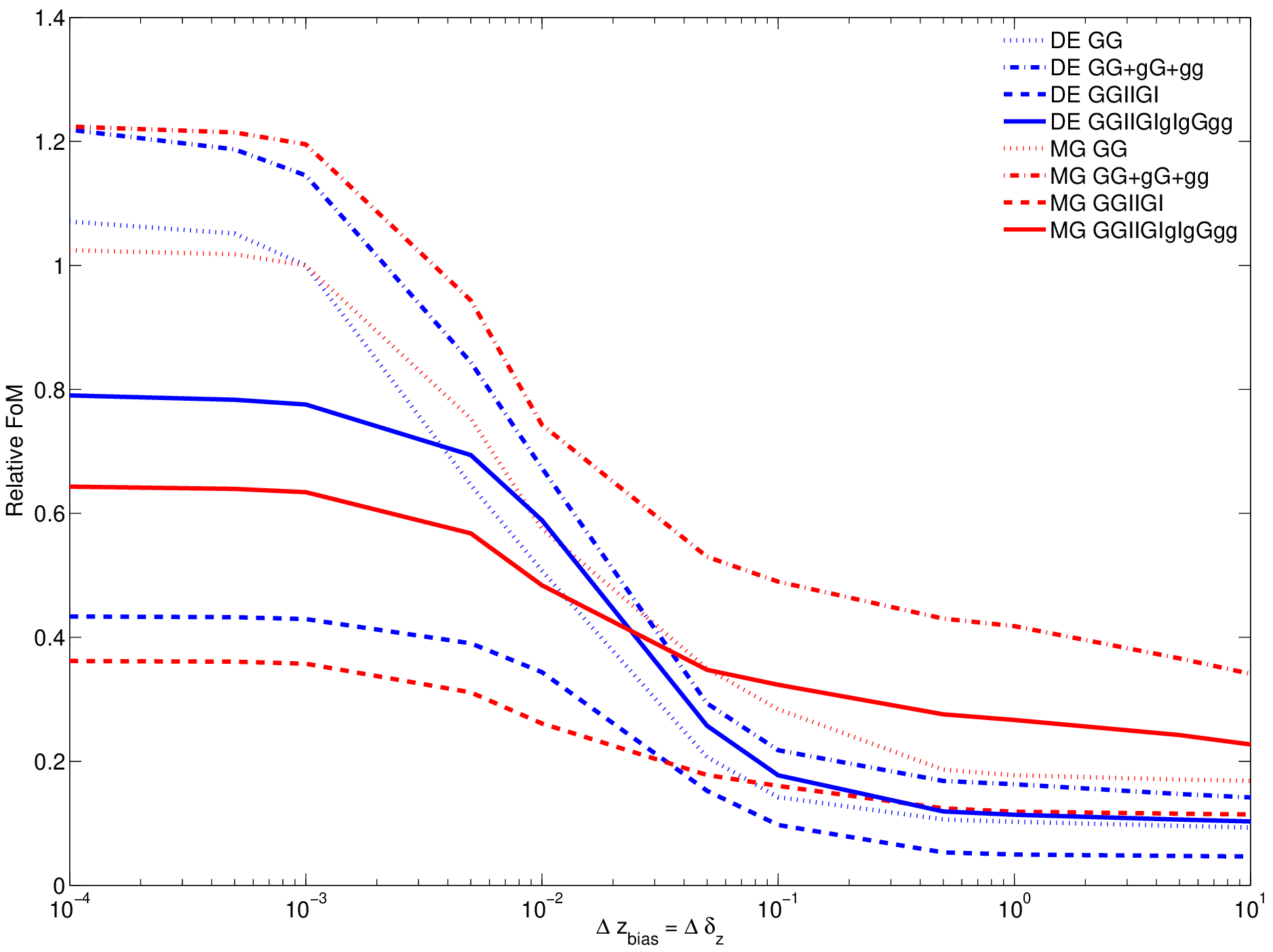}
\caption{Figure showing the variation with the priors on the bias on median redshift, $z_{bias}$, of 30 redshift bins spanning the survey redshift range due to the effect of photometric redshift uncertainties and the photometric redshift statistical dispersion, $\delta_z$. All other survey parameters are fixed at their fiducial values. Results are for a Stage IV survey FoM for dark energy (DE) parameters $w_0,w_a$ [blue lines], and modified gravity (MG) parameters $Q_0, Q_0(1+R_0)/2$ [red lines], as described in (\ref{DEFoM}) and (\ref{MGFoM}), relative to a `baseline' FoM [black cross].
As shown in the key in the figure, shear-shear correlations, `$\epsilon\epsilon$' , and combined shear and galaxy position information including cross-correlations '$\epsilon\epsilon$+$n\epsilon$+$nn$' are shown
when uncertainities in the IA model have been included and excluded.}
\label{fig:zbias_beta}
  \end{flushleft}
\end{figure}

To test this assumption we extend our photometric redshift model- breaking the redshift range 0-3 into 30 bins in z of width 0.1,
following \citet{mahh06}. Note that these bins are used only to parameterise our photo-z model, we continue to use the standard 10 bins for shear tomography. We allow $\delta_z$ to vary independently in each of these z-bins and introduce a new parameter $z_{\textrm{bias}}$, the amount by which the the photometric distribution of redshifts is offset from the true redshift. $z_{\textrm{bias}}$ is allowed to vary independently in each of our 30 z-bins. Each $\delta_z$ has the standard fiducial value of 0.05 and each $z_{\textrm{bias}}$ has fiducial value zero. We have enlarged our Fisher Matrix by 60 new free parameters, all of which are marginalised over in the results given in Fig.~\ref{fig:zbias_beta}.

We assume that the Gaussian prior assigned to each of the 60 free parameters is the same and we allow the prior to vary from a width of 0.0001 to 10. Clearly the constraining power a given survey achieves decreases as we reduce our amount of prior knowledge about the redshift distribution.

Apart from the expected decrease with wider priors we see two distinct regimes- relatively flat FoMs below prior ~ 0.001 and above 1 with a transition regime in between.
It is reassuring to note that we cannot achieve arbitrarily large FoMs with tighter redshift priors- this is connected to the previously noted ``smearing'' of redshift information in the shear angular power spectrum.
At large values of the prior width we reach a ``self-calibration'' regime in which the information in the survey is sufficient to constrain the unknown parameters \citep{Huterer:2005ez}.

When IAs are ignored, the decrease in constraining power is more pronounced. Mostly the lines start higher for narrow priors and converge to a similar low FoM for wide priors. It is probable that this behaviour is an artefact of there being fewer nuisance paramters when IAs are ignored. This produces strong constraints in the narrow prior regime, while for wide priors redshift uncertainty becomes the dominant source of error and probe combinations with/without IAs are more similar in terms of constraining power.

What is also clear is that the DE FoM falls off much more steeply over the prior range than the MG FoM. This is true for all probe combinations, with and without IAs but is most spectacular for the lines which include IAs, here there is actually a cross-over, with the MG FoM ending higher than the DE FoM at prior = 10. For $\epsilon\epsilon$, including IAs, we observe a a factor of 10 decrease in DE FoM but a factor of 3 decrease in MG FoM. This trend is repeated when all data is used ($\epsilon\epsilon+n\epsilon+nn$). We expect the dark energy parameter $w_a$ to be particularly sensitive to incomplete redshift information because it describes the time evolution of the dark energy equation of state. By contrast, the modified gravity parameters are more robust to poor redshift knowledge because they are constant with redshift and the redshift evolution of the modified gravity functions is fixed by our choice of redshift power law $s$.

We can relate the required prior on redshift parameters to the number of spectroscopic quality galaxy redshifts needed to calibrate our photometric sample as given by \citet{mahh06}:
\begin{equation}
N_{spec} = \frac{2\delta^{2}_{z}(1+z)^2}{\Delta^{2}_{\sigma_{z}}}.
\end{equation}

For $\epsilon\epsilon$, ignoring IAs, a prior of 1$\times 10^{-3}$ is sufficient to recover 90\% of the peak FoM for either DE or MG. With a photometric scatter of $\sigma_z = \delta_{z}(1+z) = 0.05(1+z)$ and a median redshift of $z_m = 0.9/\sqrt{2}$, this translates into 1.34$\times 10^4$ spectroscopic galaxies per redshift bin. Including all galaxy position information $\epsilon\epsilon + n\epsilon + nn$ and IAs reduces the required prior to $\sim 1\times 10^{-2}$, or
of order 100 spectroscopic galaxies per bin. This represents a factor of 100 reduction in required numbers compared to the standard result. However, note that there are other factors which must be taken into consideration when determining the number of galaxies per redshift bin, especially cosmic variance \citep[e.g. see the discussion in][]{VanWaerbeke:2006qt,ishakh05,bordoloiea08,cunha11}.

\begin{figure}
  \begin{flushleft}
    \centering
       \includegraphics[width=3.5in,height=4in]{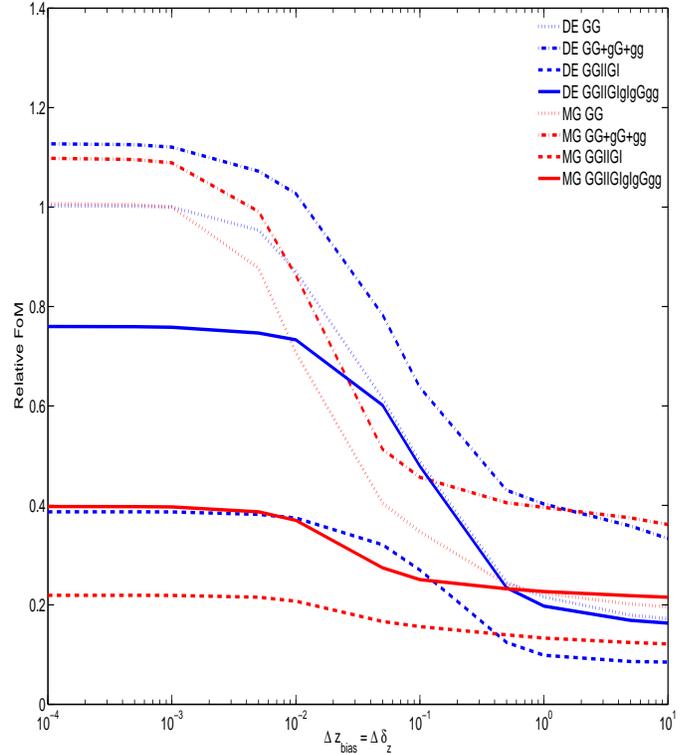}
\caption{As in figure \ref{fig:zbias_beta}, but for a Stage III survey specification.
}
\label{fig:zbias_beta_DES}
  \end{flushleft}
\end{figure}

Fig.~\ref{fig:zbias_beta_DES} shows similar results for the fiducial stage III survey. Qualitatively the results are similar to the Euclid survey. In this case, with a photometric scatter of $\sigma_z = 0.07(1+z)$ and a median redshift of $z_m = 0.8/\sqrt{2}$, $\epsilon\epsilon$ without IAs requires a prior of $\sim 1\times 10^{-2}$ which translates into around 100 galaxies per bin. Again $\epsilon\epsilon + n\epsilon + nn$ with IAs requires a factor of 100 fewer galaxies, a substantial saving. Interestingly, while the MG constraints produce the same saving in spectroscopic galaxies required when IAs and all probes are included, the absolute values are a factor of 2 tighter in prior terms (i.e. a factor of 4 more spectroscopic galaxies are required).

\section{Conclusions}
\label{sec:conclusions}

Previous work \citep{joachimi_bridle_2009} has shown how constraints from cosmic shear are significantly degraded by the inclusion of IA systematics, using a comprehensive grid of k- and z-dependent nuisance parameters. The impact of IAs can be mitigated by the inclusion of galaxy position correlations and shear-position cross-correlations using only the photometric information gathered as part of the shear survey.

In a related paper, LBKB, we confirmed and extended these results showing that using the full $\epsilon\epsilon+n\epsilon+nn$ information with up to
36 nuisance parameters can regain the constraining power of a naive $\epsilon\epsilon$ analysis which ignores IAs and that this is true when constraining deviations from GR as well as dark energy. In this paper we confirm that these results are robust across a wide range of survey parameters, encompassing fiducial stage III and stage IV cosmic shear surveys.

Increasing survey area improves both DE and MG FoMs, as one expects, whether we ignore or include intrinsic alignments. Previous work, which did not include the CMB, found a linear variation of FoM with survey area even when IAs were included.  We have shown, however, that if one includes priors from upcoming CMB experiments that  this dependence is weakened to vary roughly as a power law with an exponent of 0.4. The inclusion of IAs slightly weakens the relationship between FoM and survey area.

More realistically, survey time is a constant so increased area must be traded off against depth.
We find the dependence on area significantly flatter when IAs are included. This is to be expected because IAs dominate over the cosmic shear signal at low redshift and we effectively marginalise out information from this regime. On increasing the area, and thus decreasing the depth to lower redshift, an increasing proportion of the signal is lost by the marginalisation.
The effect is more pronounced for a stage-III survey than a stage-IV survey, which is expected because the depth is very shallow indeed if it has to cover half of the sky.

The quality of photometric redshift information is crucial to the success of future cosmic shear missions. We show that the inclusion of IAs and the full use of galaxy position information requires more accurate redshift information than a naive cosmic shear only analysis. GG-only retains 97\% of its constraining power for perfect redshift knowledge with a photometric redshift scatter of $\delta_z = 0.05$, the fiducial value for a stage-IV survey. By contrast the combined $\epsilon\epsilon + n\epsilon + nn$ including IAs requires $\delta_z = 0.015$ to retain the same proportion. Similarly the characteristic plateau behaviour, where WGL constraining power does not improve over a tomographic resolution of $\sim 4-5$ bins is pushed back on the inclusion of IAs and galaxy position information to $\sim 10-12$ redshift bins. This is understandable as the standard cosmic shear redshift dependence, modulated by a broad lensing kernel for each redshift bin, has been joined by galaxy position correlations which depend on the narrower redshift distribution function and the attempt to constrain IAs signals, each with their own redshift dependence. Taken together these place increased importance on accurate redshift knowledge.

We have presented results showing the effect of varying our degree of prior knowledge of an extended parameterisation of redshift information, allowing 60 free parameters to be shared between redshift dispersion and redshift bias.
Around $\Delta z_{\textrm{bias}} = \Delta \delta_z = 0.1$ the probes enter a self-calibration regime and the FoMs flatten off. The DE FoMs fall off much more steeply and relatively lower than their MG counterparts, highlighting the sensitivity of the DE equation of state, particularly $w_a$, to incomplete redshift information. By contrast the MG parameters have a fixed redshift dependence in this work.

This work assumed a relatively simple model for deviations from GR. We vary two MG parameters, each with the same fixed redshift dependence. In the future we hope to extend these results to take into account a MG gravity parameterisation which could vary more generally as a function of scale and redshift. In contrast our fiducial parameterisation of galaxy bias and IA nuisance parameters is robust, containing over 100 nuisance parameters. With a greater understanding of the physics of galaxy formation mechanism responsible for IAs it may be possible to motivate a more targeted, simpler parameterisation.

\section*{Acknowledgements}

We thank the Aspen Center for Physics for support and for hosting two coincident workshops on ``Wide-Fast-Deep Surveys: New Astrophysics Frontier" and  ``Testing General Relativity in the Cosmos" in 2009 where this work was conceived. The authors would like to thank the Kavli Royal Society International Centre for hosting the ``Testing general relativity with cosmology" workshop in 2011 that supported fruitful discussion and collaboration.

We thank Filipe Abdalla, Adam Amara, David Bacon, Scott Dodelson, Ole Host,  Martin Kilbinger, Andrew Jaffe,  Bhuvnesh Jain, Benjamin Joachimi, Ofer Lahav, Rachel Mandelbaum, Anais Rassat, Shaun Thomas, Alexandre Refregier and Jochen Weller for helpful discussions.

We are very grateful to Josh Frieman for spotting an error in equation 19 of the first version, and to Barney Rowe for discussing with us the correction.

RB's and IL's research is supported by NSF CAREER grant AST0844825, NSF grant PHY0968820, NASA Astrophysics Theory Program grants NNX08AH27G and NNX11AI95G and by Research Corporation.
SLB thanks the Royal Society for support in the form of a University Research Fellowship
and acknowledges support from European Research Council in the form of a Starting Grant with number 240672.
\bsp

\bibliographystyle{mn2e}
\bibliography{bibliography_DK}

\begin{thebibliography}{}

\bibitem[\protect\citeauthoryear{{Albrecht}, {Bernstein}, {Cahn}, {Freedman},
  {Hewitt}, {Hu}, {Huth}, {Kamionkowski}, {Kolb}, {Knox}, {Mather}, {Staggs} \&
  {Suntzeff}}{{Albrecht} et~al.}{2006}]{detf}
{Albrecht} A.,  {Bernstein} G.,  {Cahn} R.,  {Freedman} W.~L.,  {Hewitt} J.,
  {Hu} W.,  {Huth} J.,  {Kamionkowski} M.,  {Kolb} E.~W.,  {Knox} L.,  {Mather}
  J.~C.,  {Staggs} S.,    {Suntzeff} N.~B.,  2006, ArXiv Astrophysics e-prints

\bibitem[\protect\citeauthoryear{{Amara} \& {Refregier}}{{Amara} \&
  {Refregier}}{2006}]{amarar07}
{Amara} A.,  {Refregier} A.,  2006, ArXiv Astrophysics e-prints

\bibitem[\protect\citeauthoryear{Bacon, Refregier \& Ellis}{Bacon
  et~al.}{2000}]{Bacon:2000yp}
Bacon D.,  Refregier A.,    Ellis R.,  2000

\bibitem[\protect\citeauthoryear{{Baugh} \& {Efstathiou}}{{Baugh} \&
  {Efstathiou}}{1993}]{baugh_efstathiou}
{Baugh} C.~M.,  {Efstathiou} G.,  1993, \mnras, 265, 145

\bibitem[\protect\citeauthoryear{Bean \& Tangmatitham}{Bean \&
  Tangmatitham}{2010}]{Bean:2010zq}
Bean R.,  Tangmatitham M.,  2010, Phys. Rev., D81, 083534

\bibitem[\protect\citeauthoryear{{Benjamin}, {Heymans}, {Semboloni}, {Van
  Waerbeke}, {Hoekstra}, {Erben}, {Gladders}, {Hetterscheidt}, {Mellier} \&
  {Yee}}{{Benjamin} et~al.}{2007}]{benjaminea07}
{Benjamin} J.,  {Heymans} C.,  {Semboloni} E.,  {Van Waerbeke} L.,  {Hoekstra}
  H.,  {Erben} T.,  {Gladders} M.~D.,  {Hetterscheidt} M.,  {Mellier} Y.,
  {Yee} H.~K.~C.,  2007, \mnras, 381, 702

\bibitem[\protect\citeauthoryear{{Bernstein}}{{Bernstein}}{2009}]{bernstein_20%
08}
{Bernstein} G.~M.,  2009, \apj, 695, 652

\bibitem[\protect\citeauthoryear{Beynon, Bacon \& Koyama}{Beynon
  et~al.}{2009}]{Beynon:2009yd}
Beynon E.,  Bacon D.~J.,    Koyama K.,  2009

\bibitem[\protect\citeauthoryear{{Blake} \& {Bridle}}{{Blake} \&
  {Bridle}}{2005}]{BlakeBridle}
{Blake} C.,  {Bridle} S.,  2005, \mnras, 363, 1329

\bibitem[\protect\citeauthoryear{{Bordoloi}, {Lilly} \& {Amara}}{{Bordoloi}
  et~al.}{2010}]{bordoloiea08}
{Bordoloi} R.,  {Lilly} S.~J.,    {Amara} A.,  2010, \mnras, 406, 881

\bibitem[\protect\citeauthoryear{{Bridle} \& {King}}{{Bridle} \&
  {King}}{2007}]{bridleandking}
{Bridle} S.,  {King} L.,  2007, New Journal of Physics, 9, 444

\bibitem[\protect\citeauthoryear{{Carroll}, {Sawicki}, {Silvestri} \&
  {Trodden}}{{Carroll} et~al.}{2006}]{carrollea_fR}
{Carroll} S.~M.,  {Sawicki} I.,  {Silvestri} A.,    {Trodden} M.,  2006, New
  Journal of Physics, 8, 323

\bibitem[\protect\citeauthoryear{{Cunha}, {Huterer}, {Busha} \&
  {Wechsler}}{{Cunha} et~al.}{2011}]{cunha11}
{Cunha} C.~E.,  {Huterer} D.,  {Busha} M.~T.,    {Wechsler} R.~H.,  2011, ArXiv
  e-prints

\bibitem[\protect\citeauthoryear{{Dunkley}, {Komatsu}, {Nolta}, {Spergel},
  {Larson}, {Hinshaw}, {Page}, {Bennett}, {Gold}, {Jarosik}, {Weiland},
  {Halpern}, {Hill}, {Kogut}, {Limon}, {Meyer}, {Tucker}, {Wollack} \&
  {Wright}}{{Dunkley} et~al.}{2009}]{dunkleyetal}
{Dunkley} J.,  {Komatsu} E.,  {Nolta} M.~R.,  {Spergel} D.~N.,  {Larson} D.,
  {Hinshaw} G.,  {Page} L.,  {Bennett} C.~L.,  {Gold} B.,  {Jarosik} N.,
  {Weiland} J.~L.,  {Halpern} M.,  {Hill} R.~S.,  {Kogut} A.,  {Limon} M.,
  {Meyer} S.~S.,  {Tucker} G.~S.,  {Wollack} E.,    {Wright} E.~L.,  2009,
  \apjs, 180, 306

\bibitem[\protect\citeauthoryear{{Dvali}, {Gabadadze} \& {Porrati}}{{Dvali}
  et~al.}{2000}]{DGP}
{Dvali} G.,  {Gabadadze} G.,    {Porrati} M.,  2000, Physics Letters B, 485,
  208

\bibitem[\protect\citeauthoryear{{Eisenstein} \& {Hu}}{{Eisenstein} \&
  {Hu}}{1998}]{eisensteinhu97}
{Eisenstein} D.~J.,  {Hu} W.,  1998, \apj, 496, 605

\bibitem[\protect\citeauthoryear{{Fu} et~al.,}{{Fu}
  et~al.}{2008}]{fuea08_mnras}
{Fu} L.,  et~al., 2008, \aap, 479, 9

\bibitem[\protect\citeauthoryear{{Hetterscheidt}, {Simon}, {Schirmer},
  {Hildebrandt}, {Schrabback}, {Erben} \& {Schneider}}{{Hetterscheidt}
  et~al.}{2007}]{hetterscheidt2007}
{Hetterscheidt} M.,  {Simon} P.,  {Schirmer} M.,  {Hildebrandt} H.,
  {Schrabback} T.,  {Erben} T.,    {Schneider} P.,  2007, \aap, 468, 859

\bibitem[\protect\citeauthoryear{{Hirata} \& {Seljak}}{{Hirata} \&
  {Seljak}}{2004}]{hiratas04}
{Hirata} C.~M.,  {Seljak} U.,  2004, \prd, 70, 063526

\bibitem[\protect\citeauthoryear{{Hoekstra}, {Yee}, {Gladders}, {Barrientos},
  {Hall} \& {Infante}}{{Hoekstra} et~al.}{2002}]{hoekstraea2002RCS}
{Hoekstra} H.,  {Yee} H.~K.~C.,  {Gladders} M.~D.,  {Barrientos} L.~F.,  {Hall}
  P.~B.,    {Infante} L.,  2002, \apj, 572, 55

\bibitem[\protect\citeauthoryear{{Hu} \& {Jain}}{{Hu} \&
  {Jain}}{2004}]{hu_jain_2004}
{Hu} W.,  {Jain} B.,  2004, \prd, 70, 043009

\bibitem[\protect\citeauthoryear{Huterer, Takada, Bernstein \& Jain}{Huterer
  et~al.}{2006}]{Huterer:2005ez}
Huterer D.,  Takada M.,  Bernstein G.,    Jain B.,  2006, Mon. Not. Roy.
  Astron. Soc., 366, 101

\bibitem[\protect\citeauthoryear{{Huterer}, {Takada}, {Bernstein} \&
  {Jain}}{{Huterer} et~al.}{2006}]{huterertbj06}
{Huterer} D.,  {Takada} M.,  {Bernstein} G.,    {Jain} B.,  2006, \mnras, 366,
  101

\bibitem[\protect\citeauthoryear{{I. {Laszlo}, R. {Bean}, D. {Kirk} and S.
  {Bridle}}}{{I. {Laszlo}, R. {Bean}, D. {Kirk} and S.
  {Bridle}}}{2011}]{MGPaper1}
{I. {Laszlo}, R. {Bean}, D. {Kirk} and S. {Bridle}} 2011, Modified Gravity from
  Cosmic Shear with Intrinsic Alignments

\bibitem[\protect\citeauthoryear{{Ishak} \& {Hirata}}{{Ishak} \&
  {Hirata}}{2005}]{ishakh05}
{Ishak} M.,  {Hirata} C.~M.,  2005, \prd, 71, 023002

\bibitem[\protect\citeauthoryear{{Jain} \& {Khoury}}{{Jain} \&
  {Khoury}}{2010}]{jain_khoury}
{Jain} B.,  {Khoury} J.,  2010, Annals of Physics, 325, 1479

\bibitem[\protect\citeauthoryear{{Joachimi} \& {Bridle}}{{Joachimi} \&
  {Bridle}}{2009}]{joachimi_bridle_2009}
{Joachimi} B.,  {Bridle} S.~L.,  2009, ArXiv e-prints

\bibitem[\protect\citeauthoryear{Kaiser, Wilson \& Luppino}{Kaiser
  et~al.}{2000}]{Kaiser:2000if}
Kaiser N.,  Wilson G.,    Luppino G.~A.,  2000

\bibitem[\protect\citeauthoryear{Laszlo \& Bean}{Laszlo \&
  Bean}{2008}]{Laszlo:2007td}
Laszlo I.,  Bean R.,  2008, Phys. Rev., D77, 024048

\bibitem[\protect\citeauthoryear{{Le F{\`e}vre}, {Mellier}, {McCracken},
  {Foucaud}, {Gwyn}, {Radovich}, {Dantel-Fort}, {Bertin}, {Moreau},
  {Cuillandre}, {Pierre}, {Le Brun}, {Mazure} \& {Tresse}}{{Le F{\`e}vre}
  et~al.}{2004}]{lefevreea04}
{Le F{\`e}vre} O.,  {Mellier} Y.,  {McCracken} H.~J.,  {Foucaud} S.,  {Gwyn}
  S.,  {Radovich} M.,  {Dantel-Fort} M.,  {Bertin} E.,  {Moreau} C.,
  {Cuillandre} J.,  {Pierre} M.,  {Le Brun} V.,  {Mazure} A.,    {Tresse} L.,
  2004, \aap, 417, 839

\bibitem[\protect\citeauthoryear{{Ma}, {Hu} \& {Huterer}}{{Ma}
  et~al.}{2006}]{mahh06}
{Ma} Z.,  {Hu} W.,    {Huterer} D.,  2006, \apj, 636, 21

\bibitem[\protect\citeauthoryear{{Massey} et~al.,}{{Massey}
  et~al.}{2007}]{massey_growth2007_mnras}
{Massey} R.,  et~al., 2007, ApJS, 172, 239

\bibitem[\protect\citeauthoryear{Peacock et~al.,}{Peacock
  et~al.}{2006}]{Peacock:2006kj}
Peacock J.~A.,  et~al., 2006

\bibitem[\protect\citeauthoryear{{Schrabback} et~al.,}{{Schrabback}
  et~al.}{2009}]{schrabbackea09}
{Schrabback} T.,  et~al., 2009, ArXiv e-prints

\bibitem[\protect\citeauthoryear{{Skordis}}{{Skordis}}{2009}]{skordis_teves}
{Skordis} C.,  2009, Classical and Quantum Gravity, 26, 143001

\bibitem[\protect\citeauthoryear{{Smail}, {Ellis} \& {Fitchett}}{{Smail}
  et~al.}{1994}]{smailef94}
{Smail} I.,  {Ellis} R.~S.,    {Fitchett} M.~J.,  1994, \mnras, 270, 245

\bibitem[\protect\citeauthoryear{{Smith}, {Peacock}, {Jenkins}, {White},
  {Frenk}, {Pearce}, {Thomas}, {Efstathiou} \& {Couchman}}{{Smith}
  et~al.}{2003}]{smithea03}
{Smith} R.~E.,  {Peacock} J.~A.,  {Jenkins} A.,  {White} S.~D.~M.,  {Frenk}
  C.~S.,  {Pearce} F.~R.,  {Thomas} P.~A.,  {Efstathiou} G.,    {Couchman}
  H.~M.~P.,  2003, \mnras, 341, 1311

\bibitem[\protect\citeauthoryear{van Waerbeke et~al.,}{van Waerbeke
  et~al.}{2000}]{van_Waerbeke:2000rm}
van Waerbeke L.,  et~al., 2000, Astron. Astrophys., 358, 30

\bibitem[\protect\citeauthoryear{Van~Waerbeke, White, Hoekstra \&
  Heymans}{Van~Waerbeke et~al.}{2006}]{VanWaerbeke:2006qt}
Van~Waerbeke L.,  White M.,  Hoekstra H.,    Heymans C.,  2006, Astropart.
  Phys., 26, 91

\bibitem[\protect\citeauthoryear{Wittman, Tyson, Kirkman, Dell'Antonio \&
  Bernstein}{Wittman et~al.}{2000}]{Wittman:2000tc}
Wittman D.~M.,  Tyson J.~A.,  Kirkman D.,  Dell'Antonio I.,    Bernstein G.,
  2000, Nature, 405, 143

\bibitem[\protect\citeauthoryear{{Zhang}}{{Zhang}}{2008}]{zhang08}
{Zhang} P.,  2008, ArXiv e-prints

\end{thebibliography}

\label{lastpage}

\end{document}